\documentclass{article}
\pdfoutput=1

\usepackage{arxiv}

\usepackage[utf8]{inputenc} 
\usepackage[T1]{fontenc}    
\usepackage{hyperref}       
\usepackage{url}            
\usepackage{booktabs}       
\usepackage{amsfonts}       
\usepackage{nicefrac}       
\usepackage{microtype}      
\usepackage{graphicx}
\usepackage{gensymb}
\usepackage{xcolor}
\usepackage{mathtools}
\usepackage{upgreek}

\title{Intermixing of Fe and Cu on the atomic scale by high-pressure torsion as revealed by DC- and AC-SQUID susceptometry \textcolor{black}{and atom probe tomography}}

\author{
  Martin St\"uckler \\
  Erich Schmid Institute of Materials Science, Austrian Academy of Sciences\\
  Jahnstra{\ss}e 12, 8700 Leoben, Austria \\
  \texttt{martin.stueckler@oeaw.ac.at} 
     \And
  Heinz Krenn \\
  Institute of Physics, University of Graz\\
  Universit\"atsplatz 5, 8010 Graz, Austria
    \And
    Philipp K{\"u}rnsteiner\\
Max-Planck Institut für Eisenforschung\\
 Max-Planck Straße 1, 40237 Düsseldorf, Germany
   \And
  Baptiste Gault \\
Max-Planck Institut für Eisenforschung\\
 Max-Planck Straße 1, 40237 Düsseldorf, Germany\\
 Department of Materials, Royal School of Mines, Imperial College\\
  Prince Consort Road, London SW7 2BP, UK
   \And
   Fr{\'e}d{\'e}ric De Geuser \\
University Grenoble Alpes, CNRS\\
 Grenoble INP, SIMaP, F-38000 Grenoble, France
\And
Lukas Weissitsch\\
  Erich Schmid Institute of Materials Science, Austrian Academy of Sciences\\
  Jahnstra{\ss}e 12, 8700 Leoben, Austria 
\And
Stefan Wurster\\
  Erich Schmid Institute of Materials Science, Austrian Academy of Sciences\\
  Jahnstra{\ss}e 12, 8700 Leoben, Austria 
\And
Reinhard Pippan\\
  Erich Schmid Institute of Materials Science, Austrian Academy of Sciences\\
  Jahnstra{\ss}e 12, 8700 Leoben, Austria 
  \And
  Andrea Bachmaier\\
  Erich Schmid Institute of Materials Science, Austrian Academy of Sciences\\
  Jahnstra{\ss}e 12, 8700 Leoben, Austria 
}

\begin{document}
\maketitle

\begin{abstract}
The capability of high-pressure torsion on the preparation of supersaturated solid solutions, consisting of Cu-14Fe (wt.\%), is studied. From microstructural investigations a steady state is obtained with nanocrystalline grains. \textcolor{black}{The as-deformed state is analyzed with atom probe tomography, revealing an enhanced solubility and the presence of Fe-rich particles.} The DC-hysteresis loop shows suppressed long range interactions in the as-deformed state and evolves towards a typical bulk hysteresis loop when annealed at 500\degree C. AC-susceptometry measurements of the as-deformed state reveal the presence of a superparamagnetic blocking peak, as well as a magnetic frustrated phase, whereas the transition of the latter follows the Almeida-Thouless line\textcolor{black}{, coinciding with the microstructural investigations by atom probe tomography}. AC-susceptometry shows that the frustrated state vanishes for annealing at 250\degree C.
\end{abstract}


\section{Introduction}
Metastable solid solutions of immiscible components provide many opportunities for modifying physical properties, making them interesting for potential applications. The components forming such solid solutions should obey several conditions: they are immiscible and do not form any stable phases at thermal equilibrium. When ensuring these conditions, a continuous tunability of the mechanical and magnetic properties is possible by simply changing the composition.
\\ Regarding magnetic properties, the binary Fe-Cu system is of particular interest, because it exhibits attractive aspects regarding cost and of course because of the well-known high magnetic moment of Fe. Several studies report on the improvement and tuning of the magnetic hysteresis by preparing such solid solutions of Fe and Cu, either by vapor deposition \cite{chien1986magnetic2} or by mechanical alloying \cite{crespo1995magnetic, jiang1994magnetic, ambrose1993magnetic}. In \cite{shen2005soft}, it was shown that mechanically alloyed solid solutions of Fe-Cu exhibit soft magnetic properties. The resulting coercivity can be modeled by random anisotropy, with grain sizes below the exchange length of Fe \cite{herzer1995soft}. As in the regime of random anisotropy, the coercivity is proportional to the sixth power of the grain size, it is evident that the resulting magnetic properties are extremely sensitive to the method of preparation.
\\Apart from soft magnetic properties, the metastable Fe-Cu system exhibits also attractive magnetoresisitive properties. For instance, the granular giant magnetoresistance (GMR) was measured on mechanically alloyed Fe-Cu \cite{xiao1992giant, yermakov1996magnetoresistance, met9111188}. Granular GMR systems require ferromagnetic particles seperated from the copper matrix rather than a continuously intermixed state, whereby the resistivity can be correlated to the Fe-particle size \cite{ennen2016giant}, again making the effect extremely sensitive to the processing method and the involved parameters.
\\To benefit from the described effects on an industrial scale, it is evident that the major focus has to be directed on the method of sample preparation and on the process parameters. Whereas vapor deposition turns out not to be feasible on larger scales, upscaling is possible for mechanical alloying, but one has to prepare a bulk sample from powder. This circumstance requires an additional process, which can also give rise to a change in the physical properties, as mentioned above. Therefore, a technique, capable of upscaling as well as to directly prepare a bulk sample is desired. This is where severe plastic deformation (SPD) \cite{valiev2000bulk} comes into play. SPD involves techniques like accumulative roll bonding (ARB), equal channel angular pressing (ECAP) and high-pressure torsion (HPT). All of these approaches have in common that the sample does not change its shape during deformation. Among these techniques, especially HPT is of great interest, since the deformation can be applied continuously. Thus, values in shear strain can be reached which are hardly accessible by other techniques of SPD, which gives rise to microstructural refinement and saturation \cite{pippan2006limits}. Some studies already dealt with microstructural characterization of HPT-processed Fe-Cu \cite{kormout2017deformation, bachmaier2012formation}, but the effects described above require a certain amount of intermixing even on an atomic scale. Therefore, the microstructural investigation has now to be extended towards nano scales, thus the capability of HPT for magnetic dilution has to be addressed. 
\\In this study, a sample consisting of Cu-14Fe (wt.\%) is prepared by HPT. The focus is on the 
\textcolor{black}{correlation of microstructural data with} the magnetic properties, by combining 
\textcolor{black}{atom probe tomography (APT) data with}
SQUID-magnetometry in DC- and AC-mode. As supersaturated states are known to be metastable in thermodynamic equilibrium, a subsequent annealing treatment is expected to lead to large changes in the physical properties. To reveal the evolution of the magnetic properties in particular, also annealed states are investigated in this study, to specifically tune desired magnetic properties such as coercivity.

\section{Experimental}
High purity powders (Fe: MaTeck 99.9\% -100 +200 mesh, Cu: AlfaAesar 99.9\% -170 +400 mesh) were mixed, with an elemental composition of 14wt.\% Fe and 86wt.\% Cu, corresping to 16at.\% Fe and 84at.\% Cu. The powder mixture was hydrostatically compacted at room temperature, with an applied pressure of 5~GPa, in Ar-atmosphere to avoid contamination. The pre-compacted samples (diameter~8~mm, thickness~$\sim$0.5~mm) were exposed to HPT at 5~GPa, for 100 turns, with an applied amount of strain $\gamma$~${\sim}$~3000 at $r$~=~3~mm. A detailed description of the HPT process and sample preparation can be found elsewhere \cite{hohenwarter2009technical, stuckler2019magnetic, stuckler2020magnetic}. Subsequent annealing treatments were performed in vacuum to avoid \textcolor{black}{oxidation} ($p$~${\leq}$~10$^{-3}$~mbar) at 150\degree C, 250\degree C and 500\degree C for 1h, followed by furnace cooling. Vickers hardness testing was carried out with a Buehler Micromet 5100 in tangential sample direction (a detailed description of the sample layout and its respective orientations is given in \cite{stuckler2019magnetic}). Analysis of the microstructural evolution was performed using a Zeiss LEO 1525 Scanning Electron Microscope (SEM) in tangential sample direction. The crystalline phases constituting the material were investigated by synchrotron X-ray diffraction experiments in transmission mode (beam energy 100~keV; beam size 0.2x0.2 mm$^2$). The beam was oriented parallel to the samples axial direction, at a radius of $r$~=~3~mm. 
\textcolor{black}{Needle-shaped specimens for APT were prepared by the standard lift-out process \cite{larson2013local} in a FEI Helios NanoLab 600i dual beam focused ion beam / scanning electron microscope (FIB/SEM) device. The final specimens were sharpened  using annular milling at 30~kV acceleration voltage followed by a low kV milling at 5~kV acceleration voltage for 2~min to remove regions severely damaged by the implantation of energetic Ga ions. APT experiments were conducted using a LEAP 5000 XS at a temperature of 35~K or 40~K in laser pulsing mode.  A pulse repetition rate of 250~kHz, a pulse energy of 25~pJ and a detection rate of 1.5\% were used. The commercial software IVAS version 3.8.2 was used to reconstruct the tip volume using a radius evolution according to the voltage curve and an initial tip radius deduced from a high resolution SEM image acquired after final tip sharpening.
}
Magnetic properties were investigated by SQUID-magnetometry (Quantum Design MPMS-XL-7) in AC- and DC-mode. Therefore, samples at $r$~${\geq}$~2~mm were cut out, with the external magnetic field pointing in axial orientation of the sample.

\section{Results and Discussion}
\subsection{Microstructural evolution as a function of temperature}
\begin{figure}
\includegraphics[width=\linewidth]{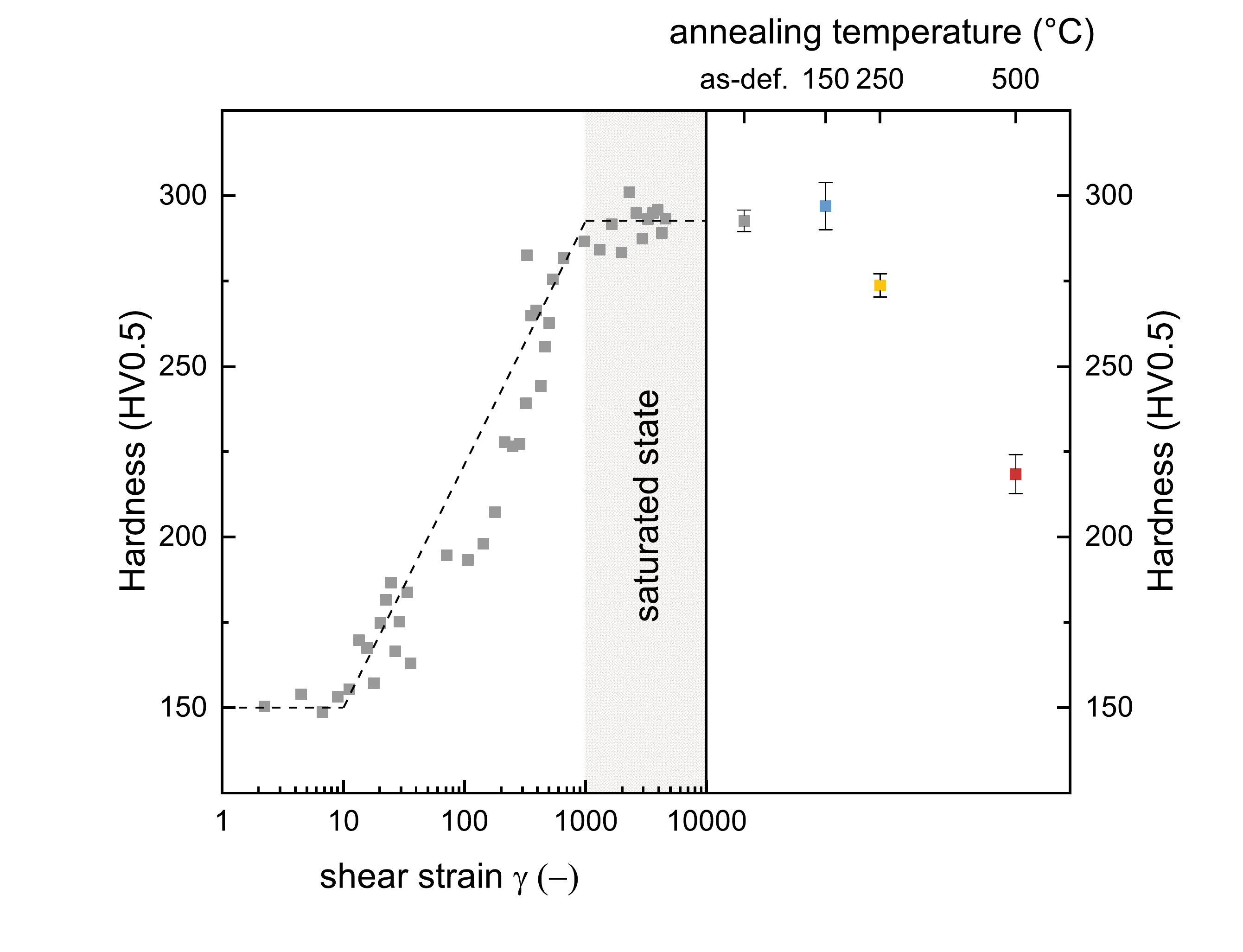}
\caption{Results obtained by Vickers hardness testing in the as-deformed and annealed states. On the left side, the hardness in the as-deformed state is plotted as a function of shear strain. A saturated state is observed at $\gamma$~${\geq}$~1000. The dotted line is a guide to the eyes. On the right side the mean hardness ($\gamma$~${\geq}$~1000) is plotted as a function of annealing temperature.}
\label{fig:hardness}
\end{figure}

\begin{figure*}
\includegraphics[width=\linewidth]{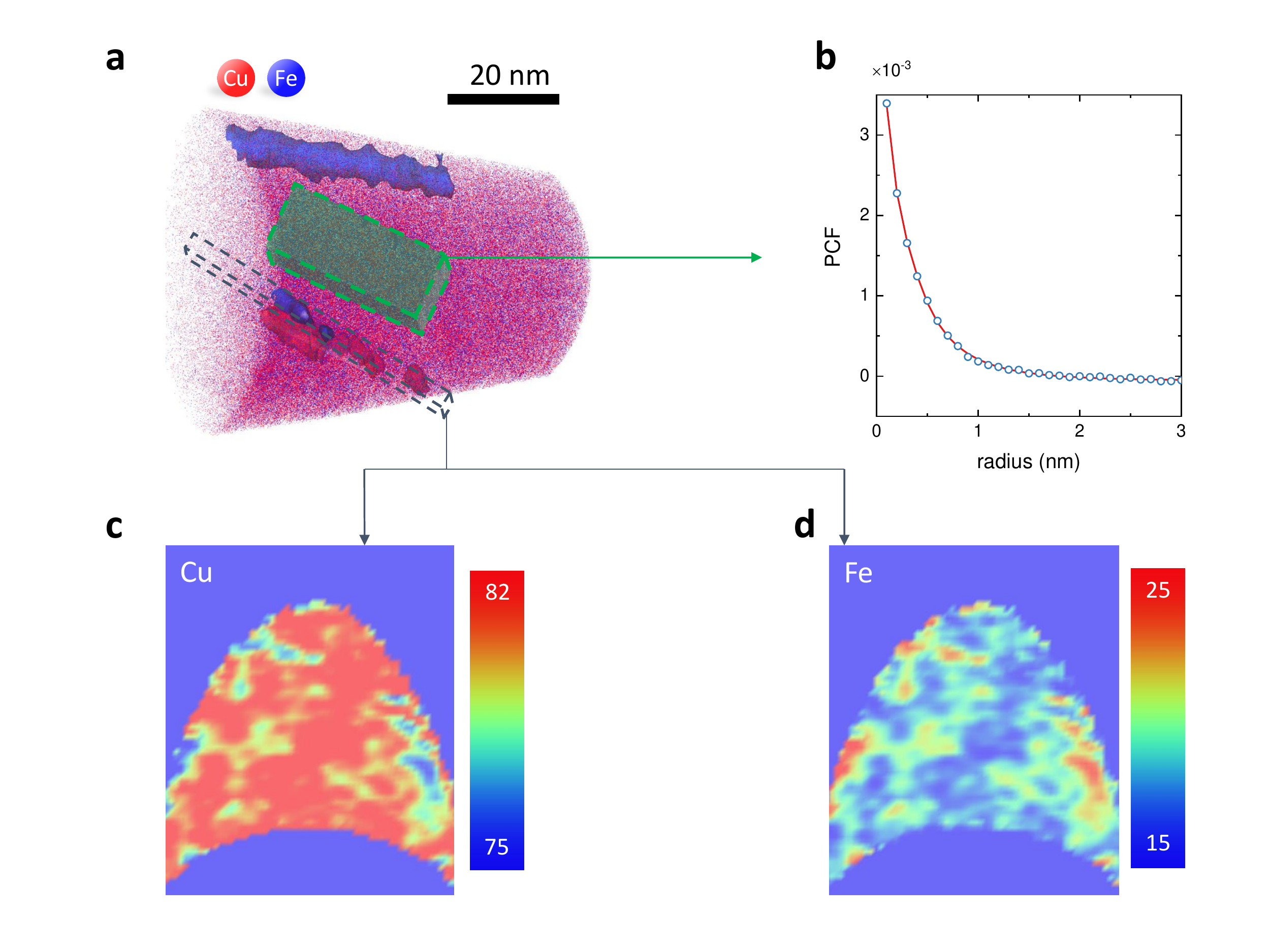}
\caption{APT reconstruction of the as-deformed state. The reconstructed volume (a) consists of Fe19.1at.\%-Cu80.8at.\%. Isosurfaces of 28.7at.\% of Fe and 63.7 at.\% of Cu are displayed. (b) shows the PCF of Fe inside the green box in (a). (c) and (d) show projections of Cu and Fe, respectively, along the grain boundary in (a). The heat maps give the concentrations in at.\%.}
\label{fig:apt2}
\end{figure*}

\begin{figure*}
\includegraphics[width=\linewidth]{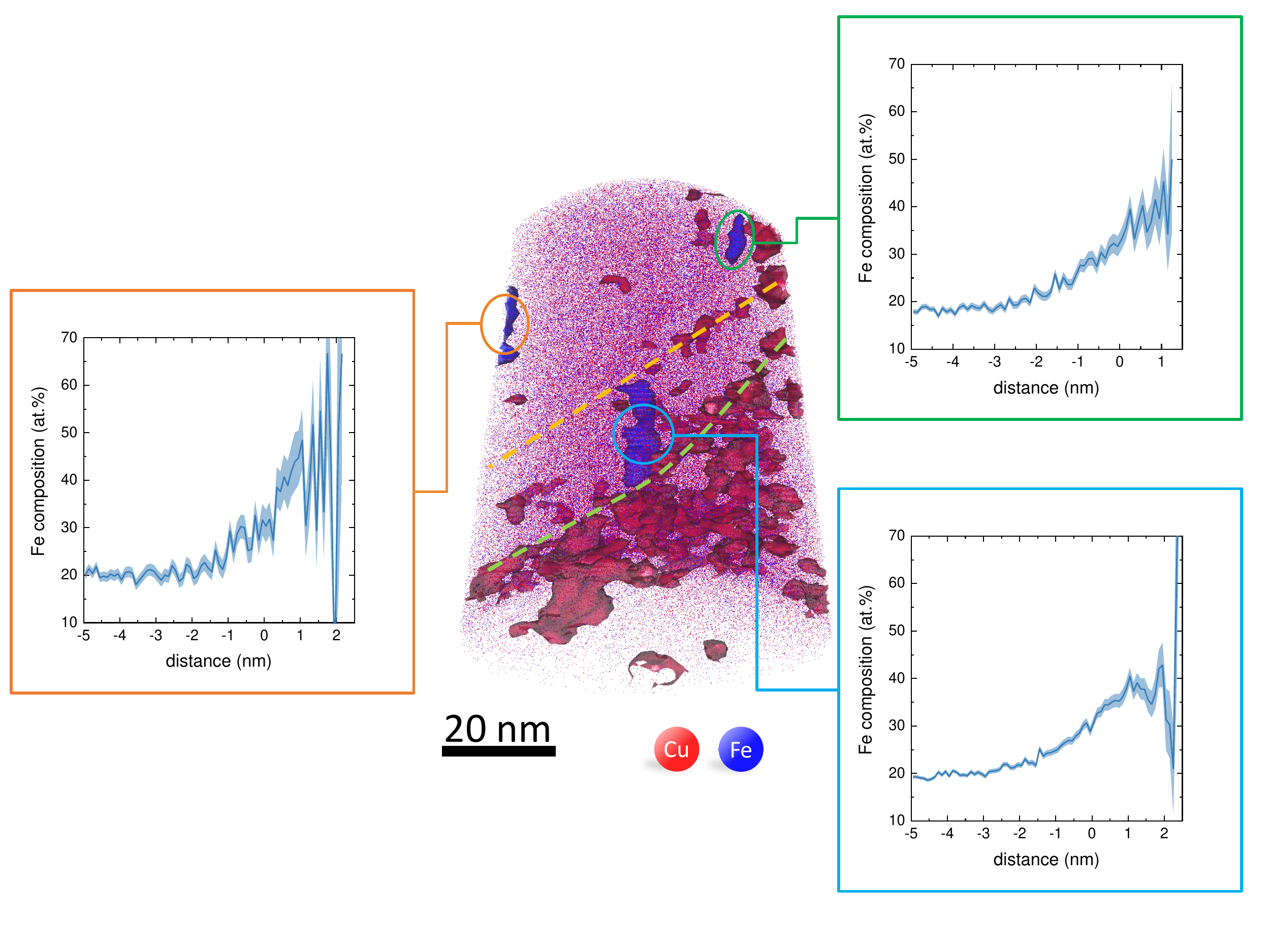}
\caption{APT reconstruction of the as-deformed state. The reconstructed volume consists of Fe19.0at.\%-Cu80.9at.\%. Isosurfaces of 28.7at.\% of Fe and 63.7 at.\% of Cu are displayed. The dashed lines mark grain boundaries, which have been detected by analyzing slices of the tomographic reconstruction, as explained in the appendix. Proxigrams of Fe-rich particles show the Fe-concentration locally going up to about 40at.\%.}
\label{fig:apt_prox}
\end{figure*}

For HPT-processing, the applied shear strain $\gamma$ increases with increasing radius [eq.~\ref{eq:strain}] \cite{valiev2000bulk}, which gives rise to microstructural evolution, e.g. grain and phase refinement \cite{pippan2006limits}.
\begin{equation}
\label{eq:strain}
\centering
 \gamma = \frac{2 \pi n r}{t}
\end{equation}
In [eq.~\ref{eq:strain}], $n$ denotes the number of turns and $t$ is the sample thickness. To confirm a saturated steady state, Vickers hardness testing in the tangential direction of the sample is performed (Fig.~\ref{fig:hardness}). The complex hardening behavior is typical for HPT-processed composites of \textcolor{black}{immiscible} components \cite{KORMOUT20172285}. A plateau at 150~HV0.5 is observed, which arises from the formation of \textcolor{black}{substructures} in the individual phases \cite{kormout2017deformation}. As this plateau is already reached at very low strains ($\gamma$~${\leq}$~1), its transition from the undeformed state is not captured experimentally. The hardness starts to increase at $\gamma$~=~10 and shows a second plateau at $\gamma$~${\geq}$~1000 and saturation. Thus, in all further measurements the sample is investigated at large radii, i.e. $r$~${\geq}$~1~mm to ensure $\gamma$~${\geq}$~1000 and therefore a saturated steady state. 
\\Annealing at 150\degree C leads to a slight increase in hardness with respect to the as-deformed state, which is characteristic for nanocrystalline materials \cite{renk2015increasing}. Upon annealing at higher temperatures, the hardness decreases, arising from grain growth, following the Hall-Petch relation \cite{hall1951proc, petch1953cleavage}.\\
\begin{figure*}
\includegraphics[width=0.24\linewidth]{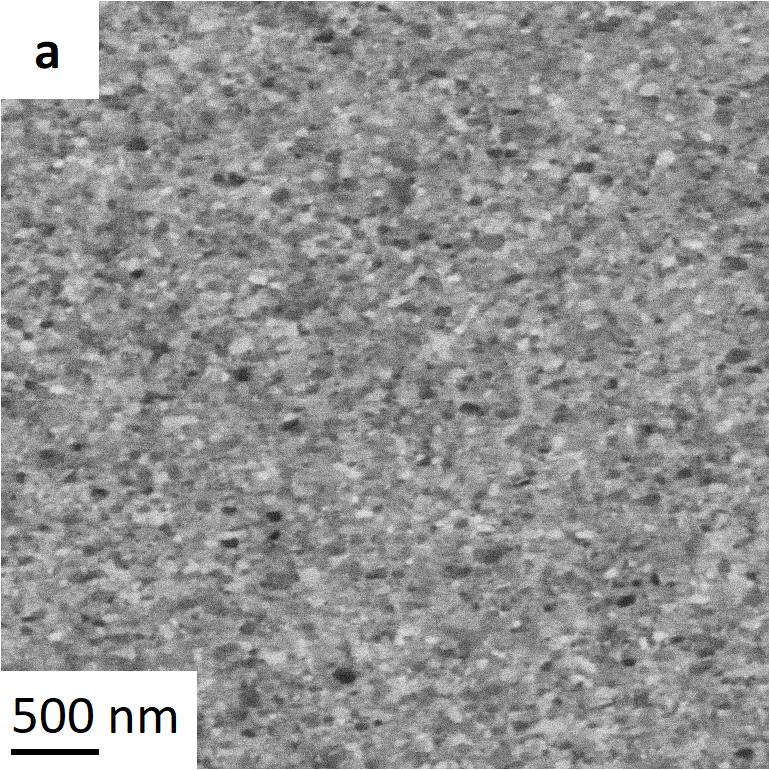}
\includegraphics[width=0.24\linewidth]{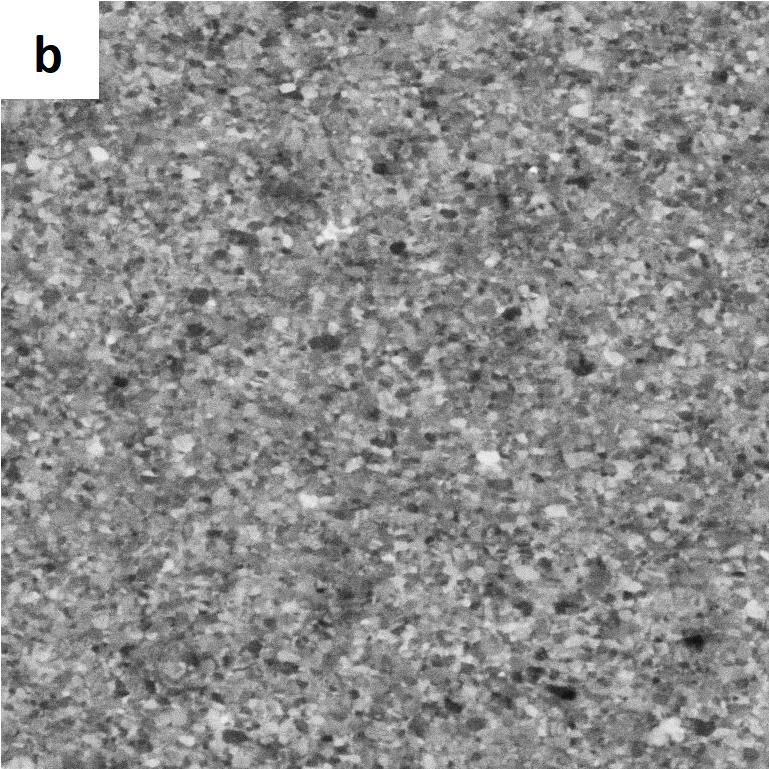}
\includegraphics[width=0.24\linewidth]{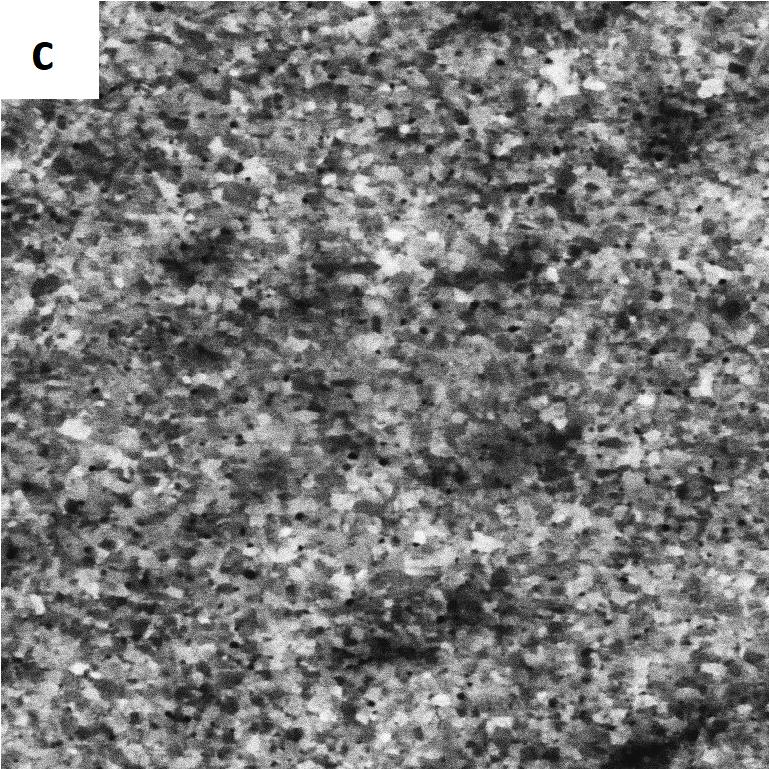}
\includegraphics[width=0.24\linewidth]{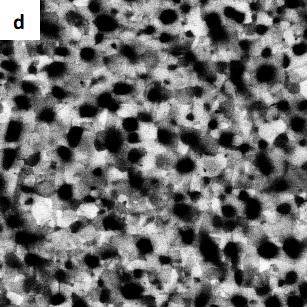}
\caption{Backscattered electron SEM images, taken in tangential direction of the sample, of the as-deformed sample (a) and samples annealed at 150\degree C (b), 250\degree C (c) and 500\degree C (d). The radial direction is parallel to the horizontal axis. The scale bar in (a) applies to all images.}
\label{fig:SEM}
\end{figure*}
Fig.~\ref{fig:SEM} shows backscattered electron micrographs that reveal the microstructural evolution of samples in the as-deformed and annealed states at $r$~${\geq}$~2~mm. The as-deformed state (Fig.~\ref{fig:SEM}a) shows a homogeneous and nanocrystalline microstructure. Neither grain growth nor demixing is observed during annealing at 150\degree C (Fig.~\ref{fig:SEM}b). The SEM image of the 250\degree C-annealed sample (Fig.~\ref{fig:SEM}c) shows the formation of dark, Fe-rich, regions and in the SEM image of the 500\degree C annealed sample (Fig.~\ref{fig:SEM}d), grain growth, as well as high phase contrast is observed.
\begin{figure}
\includegraphics[width=\linewidth]{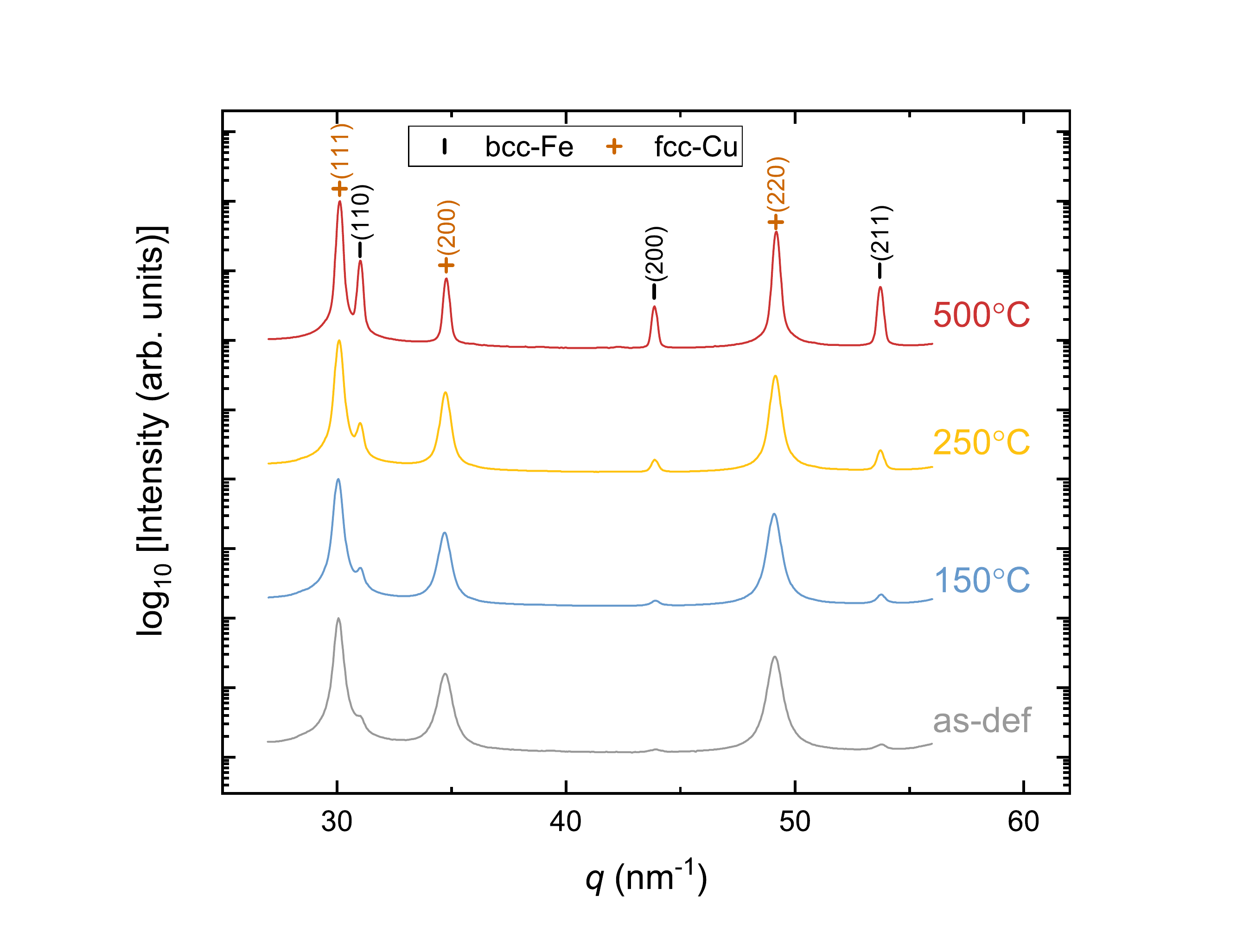}
\caption{Synchrotron-XRD diffraction patterns of as-deformed and annealed samples, measured in transmission mode.}
\label{fig:XRD}
\end{figure}
Synchrotron-XRD measurements reveal the evolution of crystalline phases during annealing (Fig.~\ref{fig:XRD}). The pattern of the as-deformed state shows dominant fcc-Cu peaks, but also weak bcc-Fe reflections can be identified. As the intense (220) Cu-peak does not overlap with another peak, the lattice constant is calculated from the peak position to $a$~=~0.362~nm, which is close to the value of pure Cu, as found in the literature ($a_{Cu,lit}$~=~0.3615~nm \cite{chiarotti19951}). The Fe-peaks are about two orders in magnitude smaller, and reveal, due to an extreme peak broadening, the presence of tiny residual Fe-particles. During annealing, the bcc-Fe peaks grow, indicating separated Fe and Cu phases for the 500\degree C-annealed sample. The peak width diminishes with annealing temperature, due to an increase in grain size.\\
As has been shown in several studies, HPT is capable to enhance the miscibility in the thermodynamically unstable Fe-Cu system \cite{quelennec2010homogeneous, lukyanov2016microstructure, kormout2017deformation}. The hardness measurements shown above (c.f. Fig.~\ref{fig:hardness}), points also to a contribution of solid solution hardening, occuring in supersaturated systems \cite{kormout2017deformation}. On the other hand, XRD-measurements (c.f. Fig.~\ref{fig:XRD}) show weak bcc-Fe peaks, contradicting a complete supersaturation of Fe in Cu. To investigate the grade of supersaturation in more detail, the distribution of Fe-atoms is analyzed by APT measurements. Two samples are investigated, whereas the overall compositions of the reconstructed data are determined to be Fe19.1at.\%-Cu80.8at.\% and Fe19.0at.\%-Cu80.9at.\%, respectively. In both cases the missing 0.1at.\% splits up into some residuals of H, C, O and Ga, whereas the latter is expected to originate from FIB milling. Fig.~\ref{fig:apt2}a show the APT reconstruction of a first specimen. Isosurfaces of 28.7at\% Fe and 63.7at.~\% Cu show inhomogeneities in both, the Fe and Cu distributions. As has been proposed in Ref.~\cite{straumal2008increase, ivanisenko2018bulk,  mazilkin2019competition}, precipitates of the minority phase are likely to occur at the grain boundary. To analyze the composition at the grain boundary, its position is determined by analyzing slices of the tomographic reconstruction perpendicular to the specimen's long axis (see appendix for more details). The grain boundary position is highlighted by the grey box in Fig.~\ref{fig:apt2}a. To analyze the composition at the grain boundary, 2D-projection maps are plotted in Fig.~\ref{fig:apt2}c and d, showing the distribution of Cu and Fe, respectively. The inhomogeneous distribution along the grain boundary shows distinct Cu- and Fe-enriched regions, with the Fe-composition locally rising to about 25at.\%. At this point it is important to mention, that in the APT reconstruction, another Fe-rich region in the upper part of the reconstructed volume is found. In contrast, no indication of a grain boundary is found in the vicinity of this particle. Fig.~\ref{fig:apt_prox} shows the reconstruction of a second specimen analyzed by APT. The dashed lines indicate the positions of the grain boundaries, which have been determined in the same way as described in the appendix. Also here, Fe-rich regions are present inside the grains. The local composition of Fe-rich regions is analyzed as shown in Fig.~\ref{fig:apt_prox}. Proxigrams \cite{gault2012atom} reveal the local concentration profiles in the proximity of the plotted isosurfaces. In Fig.\ref{fig:apt_prox}, proxigrams of three Fe-rich particles are displayed, showing the Fe-concentration locally rising to about 40at.\%.  We can therefore conclude that Fe-rich particles can be detected at grain boundaries, as well as inside the Cu-grains. Both might be responsible for the small bcc-Fe reflections in the XRD-spectrum (c.f. Fig.~\ref{fig:XRD}).
As several Fe-rich particles are found at the grain boundary as well as inside the grains, the amount of Fe, present in the grain (and furthermore being present in a supersaturated state), needs to be analyzed. Therefore, a cuboidal volume inside a grain is chosen to analyze the distribution of Fe. The overall composition of Fe inside the sketched box is $C$=18.96at.\%. We consider this volume, with Fe-composition $C$, to consist of Fe-rich precipitates with composition $C_p$ embedded in a matrix with composition $C_m$. Both parameters can be evaluated by using the DIAM-algorithm, which is based on the evaluation of the first-nearest neighbor distances \cite{Geuser2011determination}. The matrix composition is determined to be $C_m$=18.81at.\%, showing a highly supersaturated state. When assuming the presence of clusters, which consist of pure Fe (i.e. $C_p$=1) we can estimate the minimum volume fraction of Fe-rich precipitates:  $f_{V,min}=0.19$\% \cite{ZHAO2018106}. We furthermore calculate the pair correlation function (PCF), which can be used to quantify the average size of precipitates in the framework of small angle scattering \cite{COUTURIER201661, ZHAO2018106}. Fig.~\ref{fig:apt2}b shows the calculated PCF. By fitting the PCF to a function describing a lognormal assembly of spheres we can determine the average particle radius from the correlation length to be $0.35$~nm. This value should be taken only as an approximation, since particle sizes below $1$~nm should be interpreted with caution \cite{DEGEUSER2020406}, but indicates the presence of clusters consisting of several atoms. The PCF at zero distance is equal to the mean square fluctuation $(C_p-C)(C-C_m)$, but leads in the present case to $C_p~\geq~100$\%. This can likely be ascribed to an artefact of the measurement, caused by the different evaporation fields of both, the precipitate and the matrix, causing local magnification and making the precipitate to appear denser \cite{marquis_vurpillot_2008, DEGEUSER2020406}.\\
We can conclude that the analysis, restricted to a limited volume inside the grain, gives clear evidence of an enhanced solubility of Fe in Cu caused by HPT-processing \cite{quelennec2010homogeneous, lukyanov2016microstructure, kormout2017deformation}. Fe is found to be present either as solute and additionally, to a minimum fraction of $0.19$\%, as precipitate in the dimension of several atoms.

\subsection{Magnetic Properties}
\subsubsection{DC-magnetic properties}
\begin{figure}
\includegraphics[width=\linewidth]{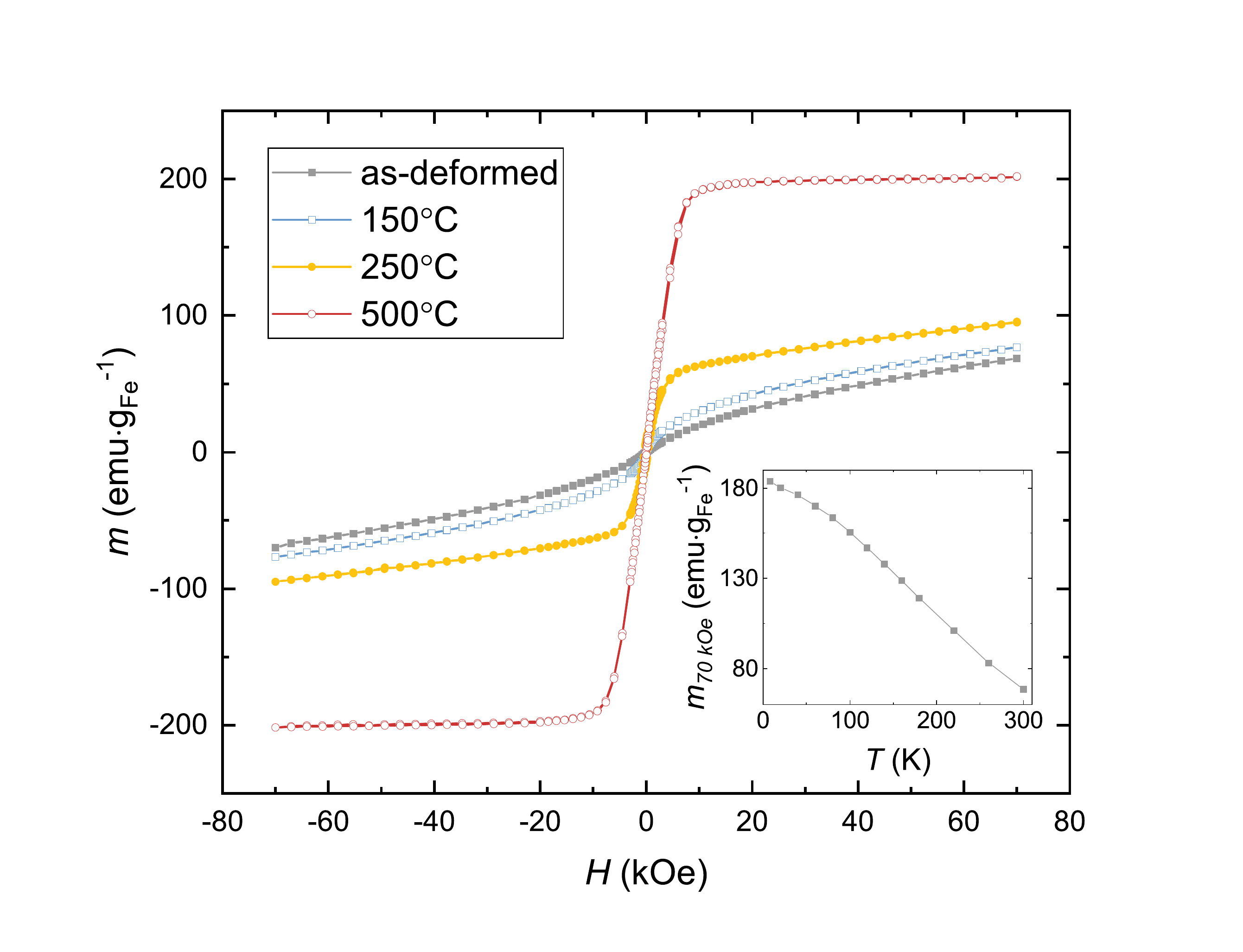}
\caption{Specific Fe-magnetization $m$ versus the applied field $H$ for Fe14wt.\%-Cu in the as-deformed and annealed states, measured at 300~K. The inlay shows the magnetic moment $m$, measured in the as-deformed state at 70~kOe, as a function of temperature $T$. The magnetic moment of bulk Fe is 218~emu$\cdot$g$^{-1}$ \cite{cullity2011introduction}.}
\label{fig:hyst}
\end{figure}
\begin{figure}
\includegraphics[width=\linewidth]{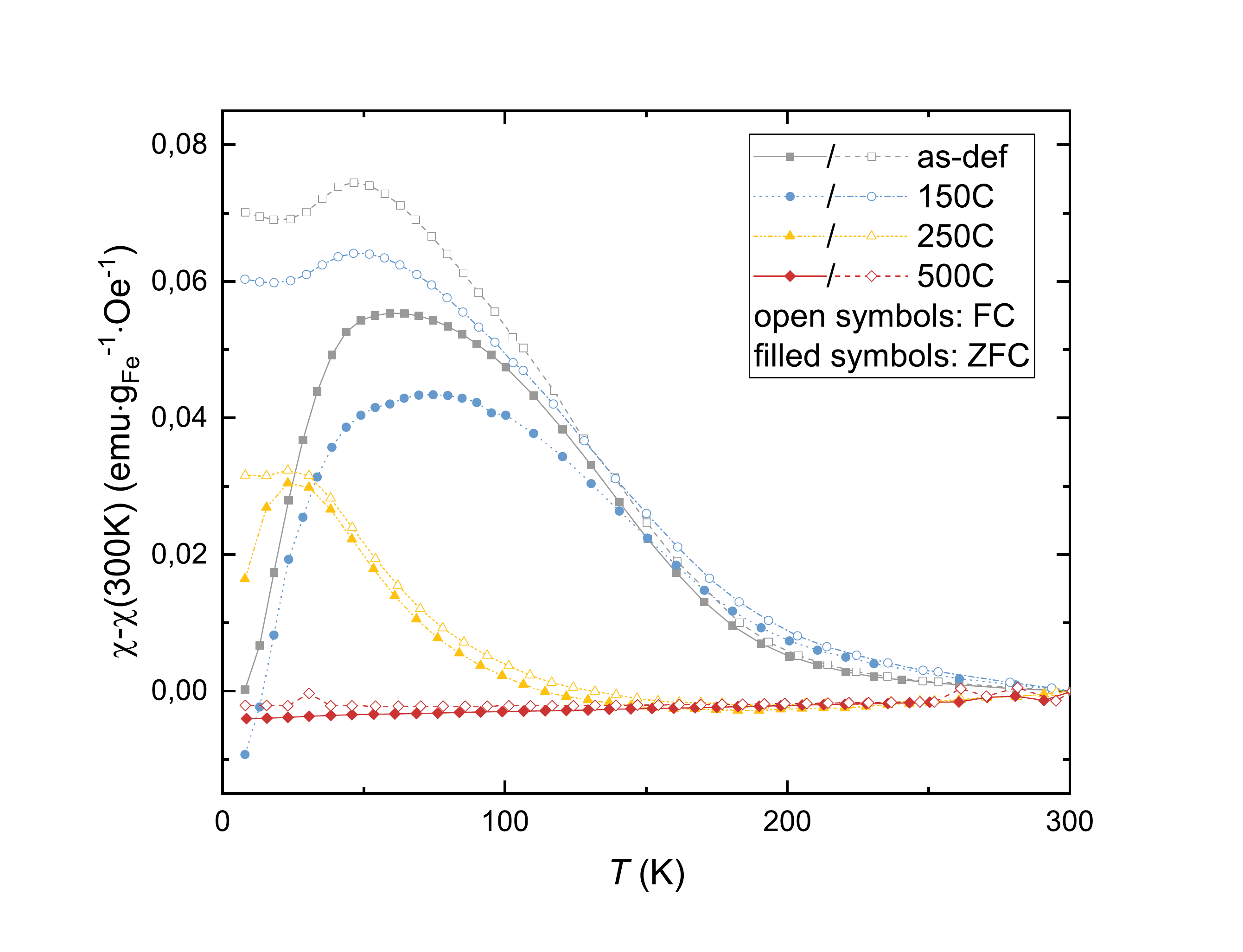}
\caption{Normalized zero-field cooling and field-cooling curves of as-deformed and annealed samples as a function of annealing temperature. In all measurements a field of $H$~=~50~Oe is applied. Open symbols represent FC-curves, filled symbols represent ZFC-curves.}
\label{fig:zfc_fc}
\end{figure}

Figure~\ref{fig:hyst} shows the DC-magnetization versus applied magnetic field (at 300~K) of samples \textcolor{black}{which were exposed to} various annealing treatments. For the as-deformed and 150\degree C-annealed sample, a significant high-field susceptibility can be identified. This behavior indicates supersaturation of Fe in the Cu-matrix, which suppresses ferromagnetic long range interaction. A non-zero high-field susceptibility is also present in the 250\degree C-annealed sample, but the slope decreases with respect to the as-deformed and 150\degree C-annealed sample, corresponding to smaller degree of supersaturation with a concomitant initiation of segregation. The 500\degree C-annealed sample shows saturation at fields higher than 1~kOe, indicating a restoration of ferromagnetic long range interaction and therefore a separation of Fe and Cu. The inlay of Fig.~\ref{fig:hyst} displays the specific magnetic moment for the as-deformed state as a function of temperature, recorded at 70~kOe. Although a strong dependence on the temperature is observed, the magnetic moment, recorded at low temperatures does not coincide with the saturation magnetization of the 500\degree C-annealed state. This deviation is expected to arise from residual paramagnetic Fe in the as-deformed state, which does not saturate even at low temperatures. 
\begin{figure}
\includegraphics[width=\linewidth]{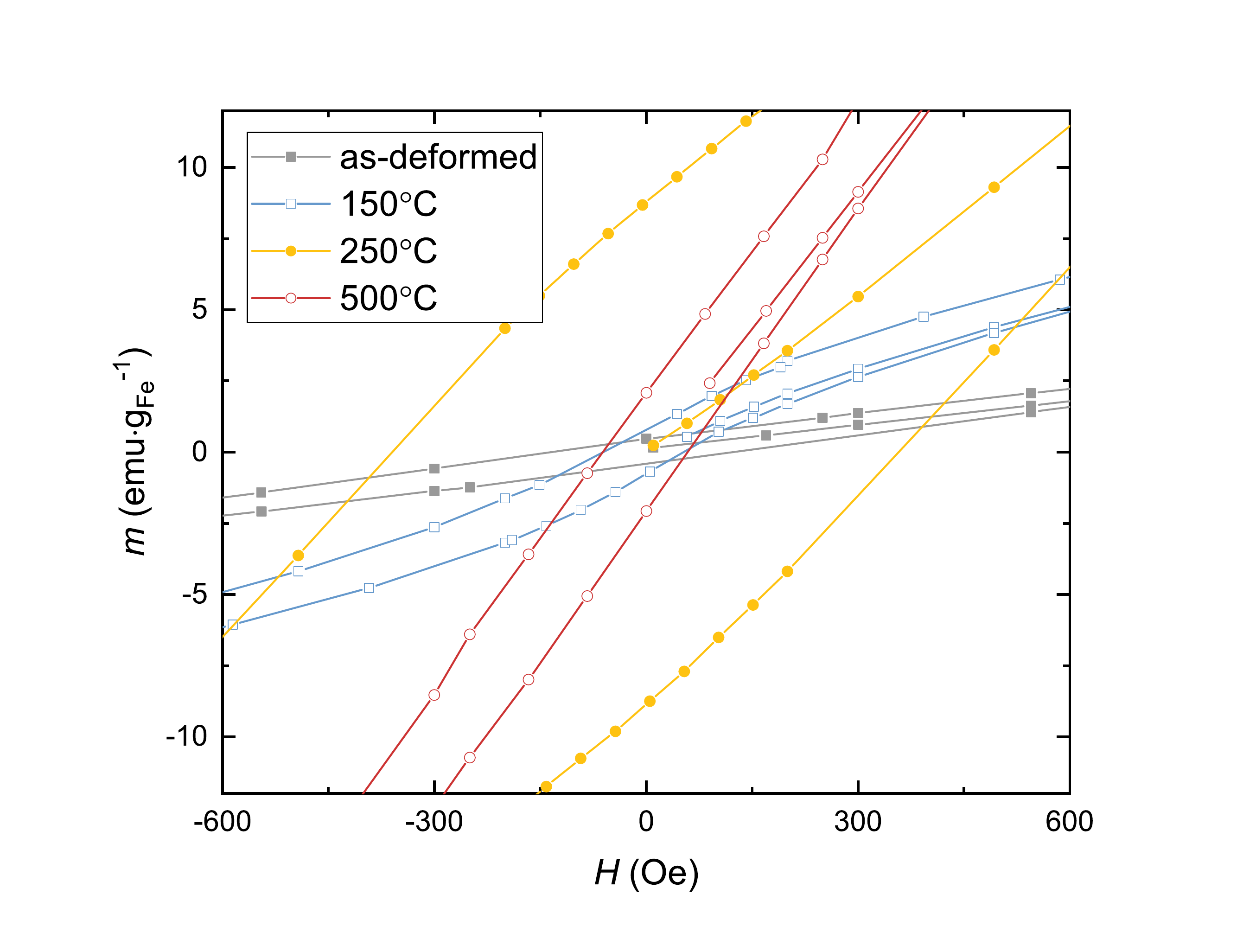}
\caption{Specific magnetization $m$ versus the applied field $H$ for Fe14wt.\%-Cu in the as-deformed and annealed states, measured at 300~K. Enlarged view of Fig.~\ref{fig:hyst}.}
\label{fig:hyst2}
\end{figure}
In Fig.~\ref{fig:hyst2} an enlarged view of the hysteresis at low magnetic fields is displayed. The coercivity is 128~Oe in the as-deformed state and decreases to 55~Oe for the 150\degree C-annealed state. HPT-deformed samples exhibit large residual stresses in the as-deformed state \cite{todt2018gradient} and therefore the diminishing coercivity is expected to arise from recovery effects, in particular a reduction in residual stresses upon slight annealing \cite{shen2005soft}. The 250\degree C-annealed state exhibits a coercivity of 358~Oe, which decreases to 60~Oe in the 500\degree C-annealed state, showing the crossover from the random anisotropy regime to the formation of multidomain particles \cite{stuckler2020magnetic}. 
\\Additionally, the temperature dependence of the low-field susceptibility was investigated. For zero-field cooling (ZFC) measurements the demagnetized sample is cooled in zero applied field. At the lowest temperature an external field of 50~Oe is applied and the magnetic moment is recorded during heating. In field-cooling (FC) temperature scans, the magnetic moment is measured during cooling in the same external field. In Fig.~\ref{fig:zfc_fc}, ZFC/FC-measurements are displayed. The curves are normalized to zero at 300~K, i.e. the recorded magnetic moment at 300~K is subtracted for reasons of comparability. The 500\degree C-annealed sample shows no splitting between the ZFC/FC-scans confirming again a reversible ferromagnetic behavior and the absence of any thermal activation below 300~K. Samples annealed below 500\degree C exhibit a splitting between the ZFC and FC curves, as well as broad peaks in the ZFC-measurements. A local maximum is observed in the FC-curves of the as-deformed state and of the 150\degree C-annealed state which is characteristic for frustrated systems, rather than for thermal activation \cite{schneeweiss2017magnetic}. Studies on the supersaturated Fe-Cu system with low amounts of Fe report the existence of a spin-glass state \cite{vedyaev1982spin, uchiyama1987spin, chien1986magnetic}.
\subsubsection{AC-magnetic properties of as-deformed state}
\begin{figure}
\includegraphics[width=\linewidth]{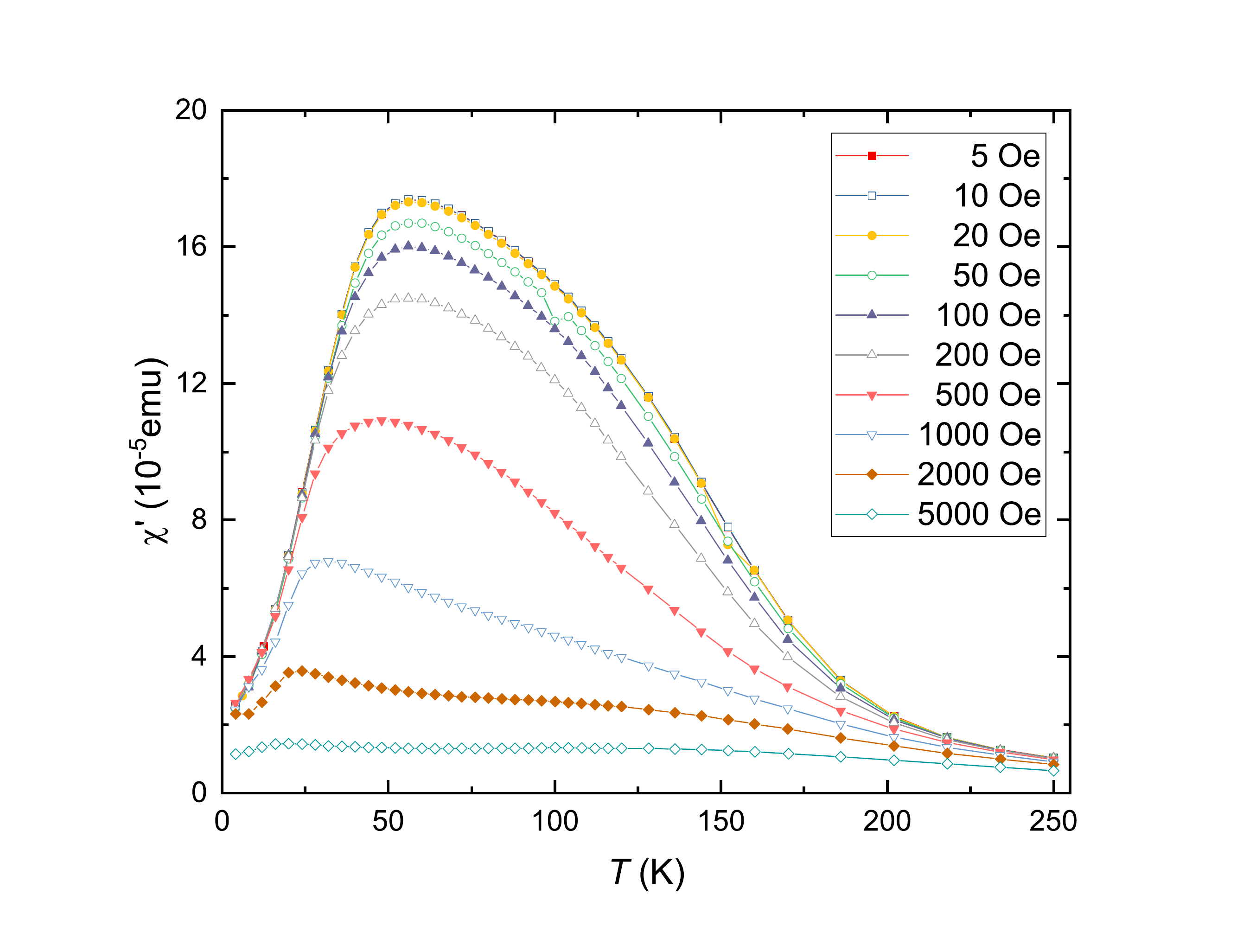}
\caption{AC-susceptibility measurement in the as-deformed state with various superimposed DC-magnetic fields. Only the in-phase component $\chi$' is shown. Measurements are taken at $f$~=~1~Hz  between 4~K and 250~K.}
\label{fig:ac_susc_H}
\end{figure}

\begin{figure}
\includegraphics[width=0.8\linewidth]{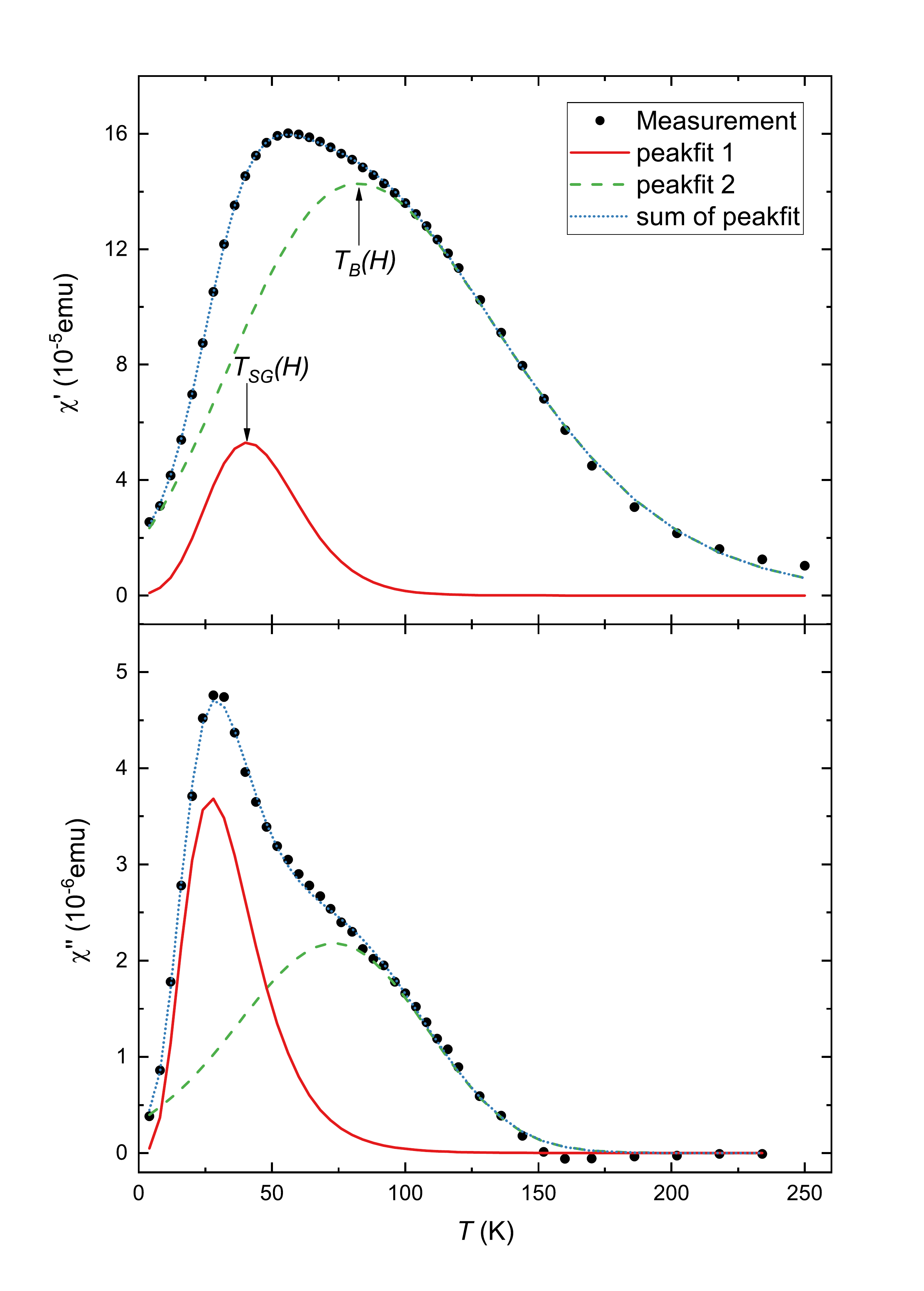}
\caption{Example of an AC-susceptibility measurement in the as-deformed state, measured between 4~K and 250~K at $f$~=~1~Hz. Here, a DC-magnetic field is applied ($H$~=~100~Oe). The driving amplitude is 5~Oe. The presence of two different peaks is clearly visible in the out-of-phase component. Analysis was carried out by fitting the peaks with two lognormal distribution functions (solid and dashed line).}
\label{fig:ac_susc}
\end{figure}

\begin{figure}
\includegraphics[width=\linewidth]{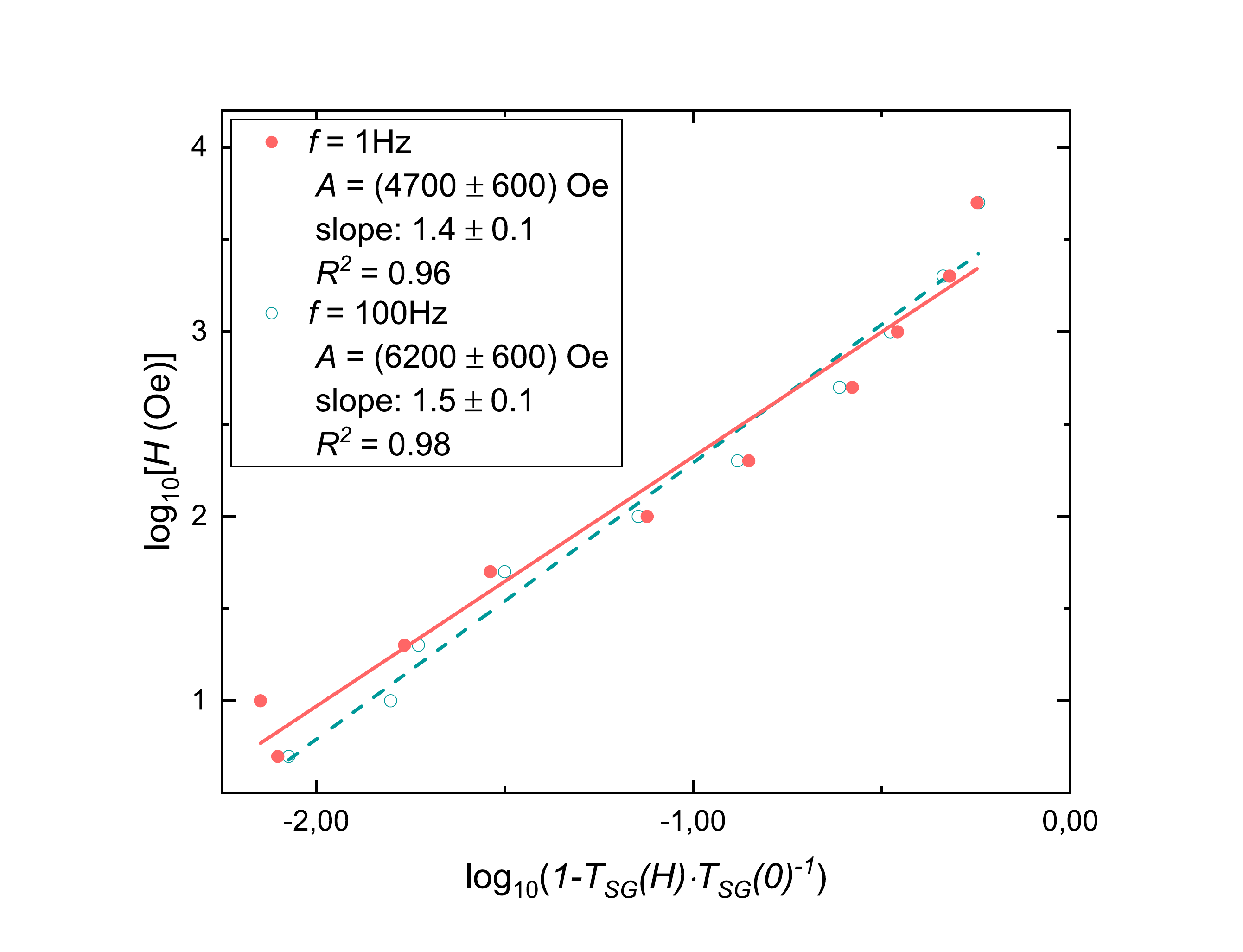}
\caption{AC-susceptibility measurement of the as-deformed state at various applied DC-fields $H$. The peak position of the lower peak from the in-phase susceptibility is evaluated. $T_{SG}(0)$ was determined from measurements at zero applied fields to 44~K ($f$=1~Hz) and 46~K ($f$~=~100~Hz). The data are in agreement with [eq.~\ref{eq:AT_line}].}
\label{fig:AT_line}
\end{figure}

\begin{figure}
\includegraphics[width=\linewidth]{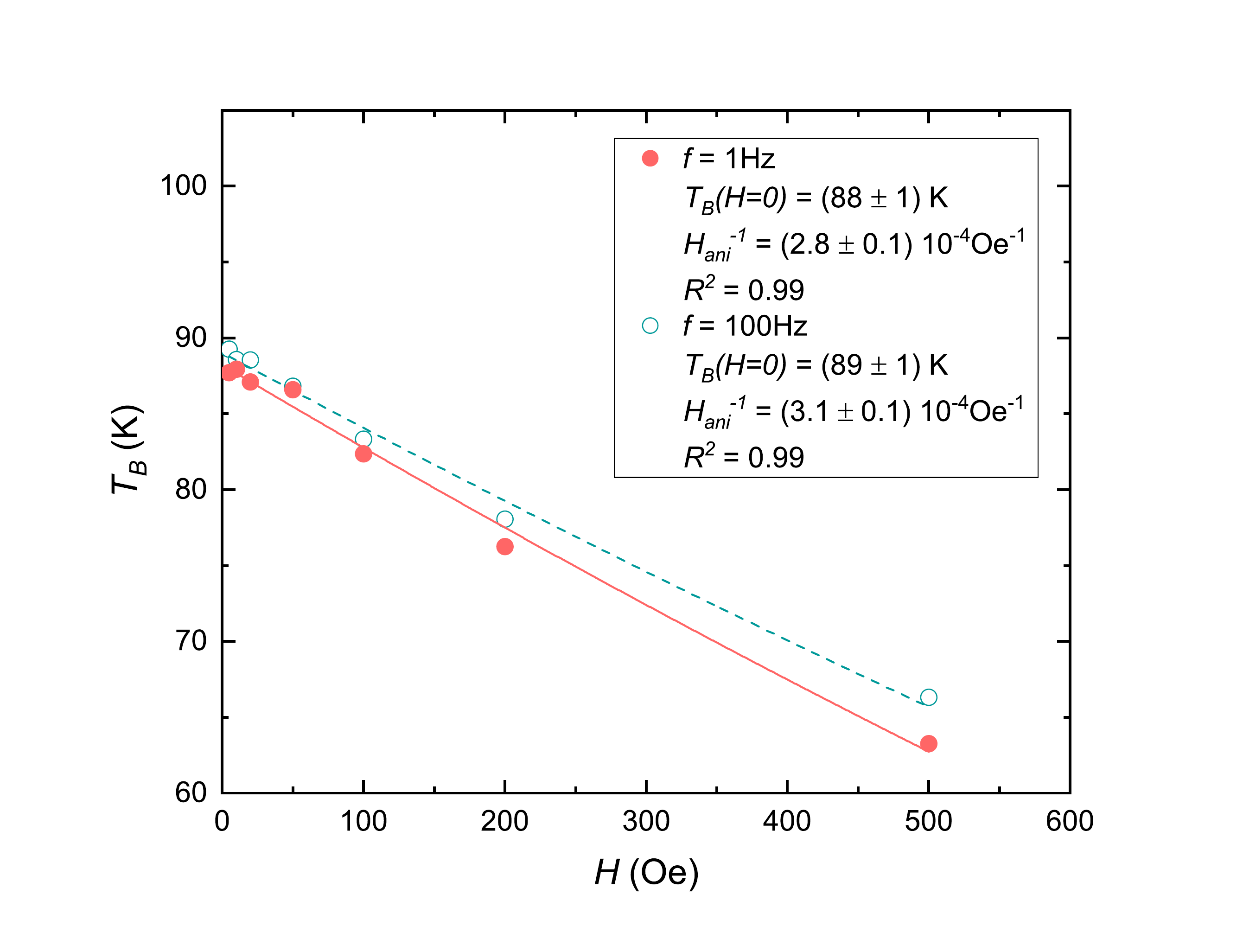}
\caption{Blocking temperature $T_B$ versus the applied DC-field $H$ for the as-deformed state. Data deduced from AC-susceptibility measurements. The lines result from least-squares fits to [eq.~\ref{eq:th_act_H}].}
\label{fig:th_act_H}
\end{figure}

To analyze the origin of the observed splittings \textcolor{black}{in} the ZFC/FC temperature scans, AC-susceptibility measurements are carried out at different frequencies $f$~=~1~Hz and $f$~=~100~Hz at an AC-amplitude of 5~Oe, with DC-fields, ranging from 5~Oe up to 5000~Oe, applied. Fig.~\ref{fig:ac_susc_H} shows the results for the measurements at $f$~=~1~Hz with various superimposed DC-magnetic fields. The observed maxima decrease in magnitude with increasing DC-magnetic field and shift towards lower temperature. Parts of the AC-susceptibilities, namely at $T$~${\geq}$~100~K, seem to decrease faster in magnitude, indicating a second peak, whose behavior at higher applied DC-fields is different. The same behavior is observed for measurements at $f$~=~100~Hz. 
Fig.~\ref{fig:ac_susc} shows the in-phase and out-of-phase components of the measurement at $f$~=~1~Hz with a superimposed magnetic field of $H$~=~100~Oe. The presence of two peaks is clearly visible in the out-of-phase component and indeed, both measurements can be fitted by two lognormal distribution functions with excellent agreement. For further analysis, the in-phase components are evaluated, as these components possess 30-times higher magnitude.
\\The low-T peak disperses sparsely with increasing DC-field and shows a small peak shift, both being characteristic features for frustrated systems. Experimental evidence of such a frustrated state can be deduced from the scaling behavior of the low-temperature peak position $T_{SG}$ for various magnetic fields. According to the theoretical Sherrington-Kirkpatrick model for spin-glass, $T_{SG}(H)$ follows the Almeida-Thouless line for not too large magnetic fields \cite{sherrington1975solvable, de1978stability}:
\begin{equation}
\label{eq:AT_line}
\centering
 H = A \left( 1 - \frac{T_{SG}(H)}{T_{SG}(0)} \right) ^{3/2}
\end{equation}
In [eq.~\ref{eq:AT_line}], $H$ denotes the applied DC-field and $A$ is a constant describing either the Heisenberg or Ising universality class of the system \cite{aruga1994experimental, sharma2010almeida}. $T_{SG}(H)$ denotes the spin-glass freezing temperature as a function of $H$, i.e. the observed peak maximum. To prove for the Almeida-Thouless line, susceptibility measurements are carried out at various DC-fields, using AC-measurements at two different frequencies ($f$~=~1~Hz, 100~Hz respectively). In Fig.~\ref{fig:AT_line} both, the applied DC-field $H$ as well as the reduced temperature, are plotted in logarithmic scale for both frequencies. From a linear fit one gets the slope in agreement with the theoretically expected value of 3/2 delivering clear evidence of the existence of a spin glass state with random dilution of the magnetic entities.
\\In contrast to the low-T peak, the high-T peak $T_{B}(H)$ disperses strongly with increasing field. This behavior is indicative to thermal activation ("magnetic blocking"). Also no proper scaling (power) law like in magnetic frustrated systems was found. For thermal activated systems anticipating mono-domain particles, the magnetic-field-dependence of the blocking temperature $T_{B}(H)$ can be described according to [eq.~\ref{eq:th_act_H1}] \cite{knobel2008superparamagnetism}.
\begin{subequations}
\label{eq:th_act_H}
\begin{equation}
\label{eq:th_act_H1}
\centering
T_{B}(H) = T_{B}(\mathbin{{H}{=}{0}}) \cdot \left( 1- \frac{H}{H_{ani}} \right)^2
\end{equation}
\begin{equation}
\label{eq:th_act_H3}
\centering
 H_{ani} = \frac{2K}{M_S}
\end{equation}
\end{subequations}
$H_{ani}$ denotes the anisotropy field, $M_S$ is the saturation magnetization and $K$ is the bulk magnetic anisotropy energy. The blocking temperature in absence of a DC-field $T_B(\mathbin{{H}{=}{0}})\mathbin{=}T_B$ obeys the Arrhenius-law [eq.~\ref{eq:th_act}].
\begin{equation}
\label{eq:th_act}
\centering
 \tau(f) = \tau_0~exp\left(\frac{K V}{k_B~T_{B}}\right)
\end{equation}
In [eq.~\ref{eq:th_act}] $\tau_0$ is the characteristic atomic precession time constant of the order $10^{-9}$~-~$10^{-12}$~s, $k_B$ is the Boltzmann constant, $T_{B}$ is the blocking temperature, $V$ is the volume of the superparamagnetic particle and $\tau(f)$ denotes the relaxation time (i.e. the inverse of the AC-fields frequency) \cite{knobel2008superparamagnetism}. Fig.~\ref{fig:th_act_H} shows the behavior of $T_B$, plotted versus the applied DC-field, up to $H$~=~500~Oe. For higher values in $H$, a proper allocation of the peak maximum is hardly possible, due to an increased broadening of the peak width. $T_B$ decreases with increasing DC-field, causing a lowering of the energy minimum for magnetization parallel to the direction of $H$. A least-mean square fit to [eq.~\ref{eq:th_act_H1}] yields results for $T_{B}(\mathbin{{H}{=}{0}})$ and $H_{ani}$. The latter can be related to the magnetic saturation $M_S$ [eq.~\ref{eq:th_act_H3}]. Assuming the magnetic anisotropy to arise from magnetocrystalline anisotropy only ($K=K_1^{Fe}=4.8\cdot10^4$~J/m$^3$ \cite{gengnagel1968temperature}), yields $M_S$~=~3.4~kG for the measurements at $f$~=~1~Hz and $M_S$~=~3.7~kG for the measurement at $f$~=~100~Hz, leading to 220~emu$\cdot$g$^{-1}$ and 239~emu$\cdot$g$^{-1}$ respectively, in good agreement with the bulk saturation magnetization of Fe \cite{cullity2011introduction}. From $T_{B}(\mathbin{{H}{=}{0}})$ an average particle size can be estimated [eq.~\ref{eq:th_act}], leading to 9~-~10~nm for both frequencies, when assuming $\tau_0$~=~10$^{-9}$~s and spherical particles.
\\To sum up, the described AC-magnetic measurements reveal the presence of two different magnetic phases in the as-deformed state: superparamagnetism can be attributed to residual Fe-particles. On the other hand, the magnetic frustrated phase is expected to arise from individual Fe-atoms diluted in the Cu-matrix, i.e. the supersaturated state.

\subsubsection{AC-magnetization: characterization of annealed states}
\begin{figure}
\includegraphics[width=\linewidth]{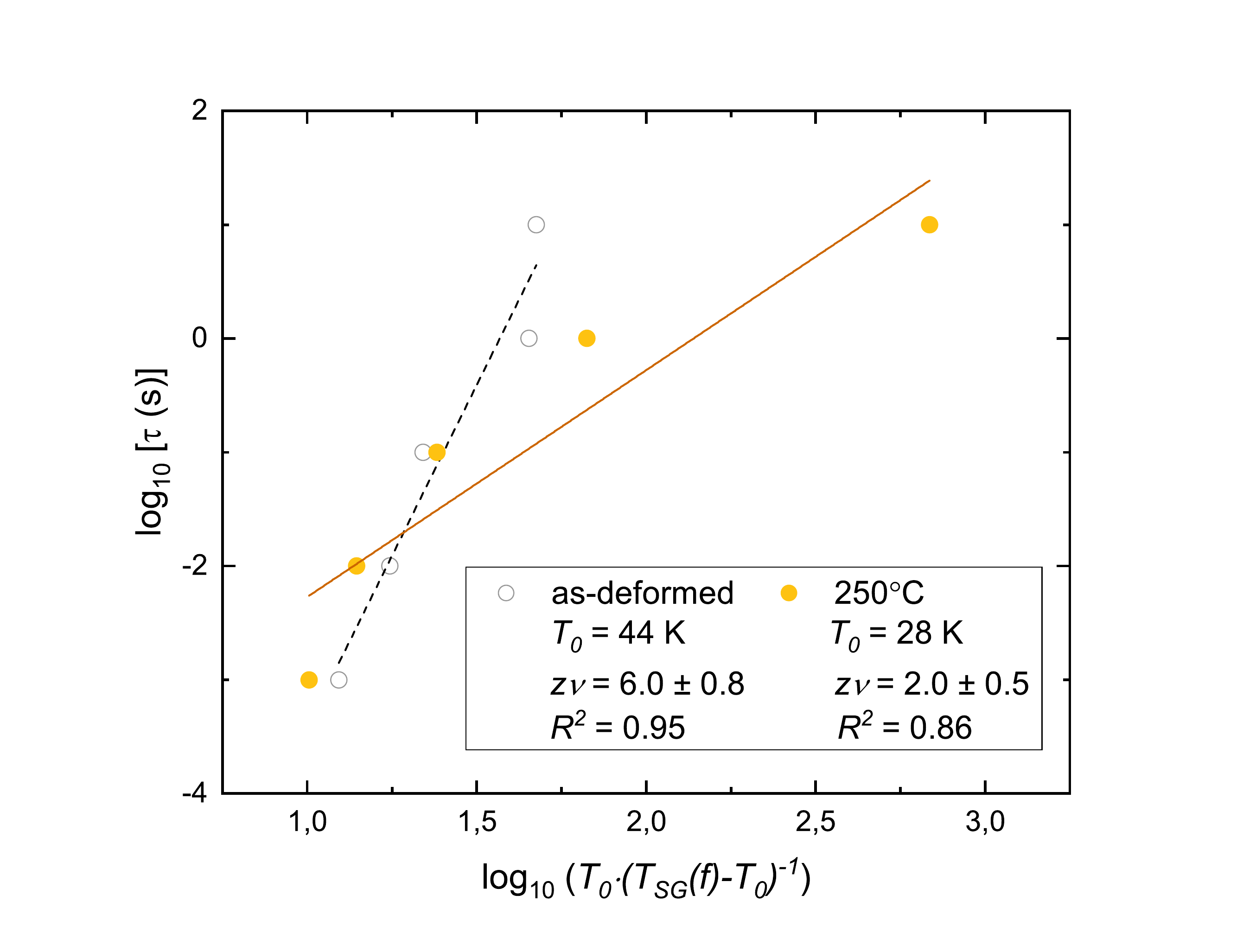}
\caption{Logarithmic plot of the dynamic scaling law [eq.~\ref{eq:dynamic_scaling}]. Data deduced from AC-susceptibility measurements for the as-deformed and 250\degree C-annealed sample. $T_0$ was determined by extrapolating the data towards $f$~=~0~Hz.}
\label{fig:dyn_sc}
\end{figure}
In contrast to the as-deformed and 150\degree C-annealed state, the FC-curve does not show a significant cusp for the 250\degree C-annealed state, but a splitting in the ZFC/FC remains (cf. Fig.~\ref{fig:zfc_fc}). To investigate its origin, AC-susceptibility measurements are performed between 4~K and 250~K for various frequencies $f$=0.1~Hz~-~1~kHz in absence of an external field $H$. As mentioned above, the as-deformed sample shows \textcolor{black}{solute}  Fe-atoms in the Cu-matrix. Therefore, the grade of \textcolor{black}{supersaturation} for the 250\degree C-annealed sample can be evaluated by testing for the dynamic scaling law [eq.~\ref{eq:dynamic_scaling}], which applies for spin-glasses \cite{balanda2013ac, zeb2016surface}.
\begin{equation}
\label{eq:dynamic_scaling}
\centering
 \tau(f)=\tau^* \left( \frac{T_0}{T_{SG}(f)-T_0} \right)^{z\nu}
\end{equation}
In [eq.~\ref{eq:dynamic_scaling}], $\tau(f)$ denotes the relaxation time, i.e. the inverse of the AC-fields frequency, $\tau^*$ is a constant, $T_0$ is the static freezing temperature and $T_{SG}(f)$ is the freezing temperature at a specific frequency $f$, i.e. temperature of the observed peak maximum in the $T$-scan. For spin glasses, typical values of $z\nu$ are 4~-~12 \cite{arauzo2015spin}. $T_0$ is obtained by extrapolating $T_{SG}(f)$ to $f$~=~0~Hz. The data are plotted in Fig.~\ref{fig:dyn_sc}. For the as-deformed sample $z\nu$~=~6.0~$\pm$~0.8, which denotes a spin-glass behavior, whereas for the 250\degree C-annealed sample the data do not scale in a proper manner. This is also reflected by the low value of fitting confidence $R^{2}$~=~0.86 for the latter case. The least-squares fit delivers a slope of 2.0~$\pm$~0.5, which does also not coincide with the expected values of $z\nu$~=~4~-~12, leading to the conclusion that the spin glass behavior is not anymore dominant in the 250\degree C-annealed state. In contrast, the results obtained for the as-deformed sample fit quite well to the model of magnetic frustration and spin-glass behavior.
\\From the large coercivity in the 250\degree C annealed state (cf. Fig.~\ref{fig:hyst2}), also the presence of larger Fe-particles is assumed in this state, which cannot be traced by AC-magnetometry \cite{stuckler2020magnetic}. Additionally, a small amount of \textcolor{black}{solute}, non-interacting, atomic iron clusters persists in (super)paramagnetic state, as the hysteresis loop does not saturate (cf. Fig.~\ref{fig:hyst}).

\subsection{RKKY-interaction as origin of the magnetic frustrated phase}
\begin{figure}
\includegraphics[width=\linewidth]{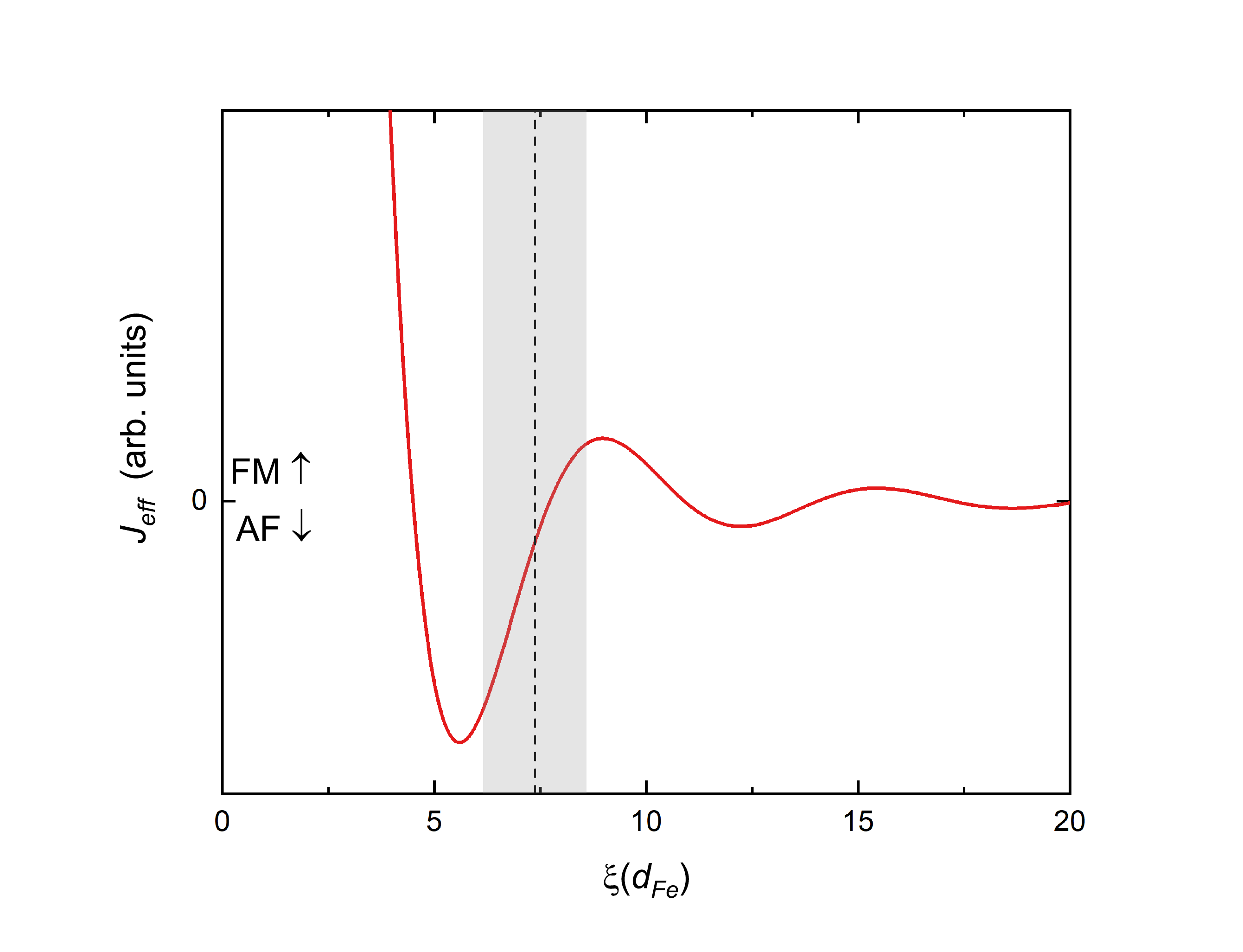}
\caption{RKKY-Interaction calculated for $k_{F,Cu}$ [eq.~\ref{eq:rkky}]. The dashed line represents $\xi$ as estimated from simulation, with the shaded area being the respective standard deviation. $\xi$ spans over a wide range, giving rise to ferromagnetic (FM) and antiferromagnetic (AF) interactions.}
\label{fig:rkky}
\end{figure}

Magnetic frustration, as obtained with AC-magnetometry, can be attributed to randomly distributed, separated Fe-atoms, coupled via RKKY-interaction \cite{binder1986spin, franz1973magnetic}. Varying nearest neighbor distances give rise to either ferromagnetic or antiferromagnetic interaction ("bond-disorder"). To estimate the interatomic Fe-Fe distances, a 50x50x50 fcc-supercell is simulated (periodic boundary conditions applied), whereas 16\% of all sites are randomly occupied, corresponding to the samples Fe-content (14 wt.\% $\equiv$ 16 at.\%). The lattice constant of the supercell is set to 0.362~nm, as obtained by synchrotron-XRD. Evaluating the interatomic distances of the supercell leads to a mean Fe-Fe distance of  $d_{Fe}$ = (0.27 $\pm$ 0.04)~nm. To calculate the RKKY-interaction [eq.~\ref{eq:rkky}] only the nearest-neighbor interaction is taken into account.
\begin{subequations}
\label{eq:rkky}
\begin{equation}
\centering
J_{eff} = J^* \frac{sin(\xi)-\xi \cdot cos(\xi)}{~\xi^4}
\end{equation}
\begin{equation}
\centering
 \xi = 2k_F \cdot d_{Fe}
\end{equation}
\end{subequations}
In [eq.~\ref{eq:rkky}], $J^*$ is the on-site exchange constant, arising from the contact-interaction between localized Fe-spins and delocalized conduction electrons of the copper matrix. $\xi$ is a dimensionless parameter of the Fermi wave vector $k_F$ and the interatomic Fe distance $d_{Fe}$. Fig.~\ref{fig:rkky} shows a plot of the RKKY-interaction, calculated for $k_F$~=~$k_{F,Cu}$, as the matrix consists of Cu. The spread of $\xi$ was deduced from the standard deviation of interatomic Fe-Fe distances as calculated above. $\xi$ spans over a wide range of $J_{eff}$, covering antiferromagnetic, as well as ferromagnetic interactions, capable to explain the magnetic frustration in the as-deformed state.\\
\textcolor{black}{
It should be stated that the presented model is a simple approach when comparing it to the the complex microstructure, which has been revealed by APT. The exact grades of supersaturation are expected to vary locally, but, as shown by APT, it can locally even exceed the overall Fe-content of 16at.\%.  Also, Fe-particles which are expected to give rise to superparamagnetism in the as-deformed state are not included. The above presented model should therefore be taken as an approximation of the real system but can sufficiently explain the origin of the magnetically frustrated phase, closely related to the enhanced grade of supersaturation caused by HPT-deformation.
}

\section{Conclusion}
A powder mixture of Fe and Cu is cold compacted and subsequently deformed with HPT until a saturated microstructure is generated. The grade of intermixing is investigated by correlating microstructural analysis with magnetometry. 
\textcolor{black}{
APT experiments reveal a complex microstructure in the as-deformed state. Apart from Fe-rich particles being present at the grain boundaries and inside the grains, the grains itself exhibit a larger solubility as in the thermodynamical equilibrium. The supersaturated grains are found to consist of solute Fe as well as tiny precipitates, being a direct evidence of the capability of SPD methods to enhance the solubility limit with respect to the thermodynamical equilibrium.
AC-susceptometry measurements  of}
the as-deformed sample \textcolor{black}{reveal} a magnetic frustrated state, as well as thermally activated behavior. The magnetic frustrated state was proven by temperature scans as a check of the dynamic scaling law and by magnetic-field shifts following the Almeida-Thouless line \textcolor{black}{ 
and can be associated to supersaturated solid solutions of Cu-Fe. The thermal activated behavior is expected to arise from residual Fe-particles.
}
\\The frustrated and thermal activated magnetic phases persist upon annealing at 150\degree C, but a diminishing coercivity points out that a reduction in residual stresses takes place. The magnetic frustrated phase vanishes during annealing at 250\degree C. Despite, larger Fe-particles give rise to an enhancement in coercivity. The 500\degree C annealed state shows a fully decomposed microstructure, with a bulk ferromagnetic behavior and the formation of multidomain particles with its typical hysteresis loop.

\section*{Acknowledgments}
This project has received funding from the European Research Council (ERC) under the European Union’s Horizon 2020 research and innovation programme (Grant No. 757333). Synchrotron measurements leading to these results have been performed at PETRA III: P07 at DESY Hamburg (Germany), a member of the Helmholtz Association (HGF). We gratefully acknowledge the assistance by Norbert Schell and Peter Knoll and appreciate the encouraged help with data processing by Florian Spieckermann and Christoph Gammer. The authors thank Manoel Kasalo for sample preparation. \textcolor{black}{We thank Uwe Tezins, Christian Broß and Andreas Sturm for their support to the FIB and APT facilities at MPIE.}

\section*{Determination of grain boundary positions}
\begin{figure*}
\includegraphics[width=\linewidth]{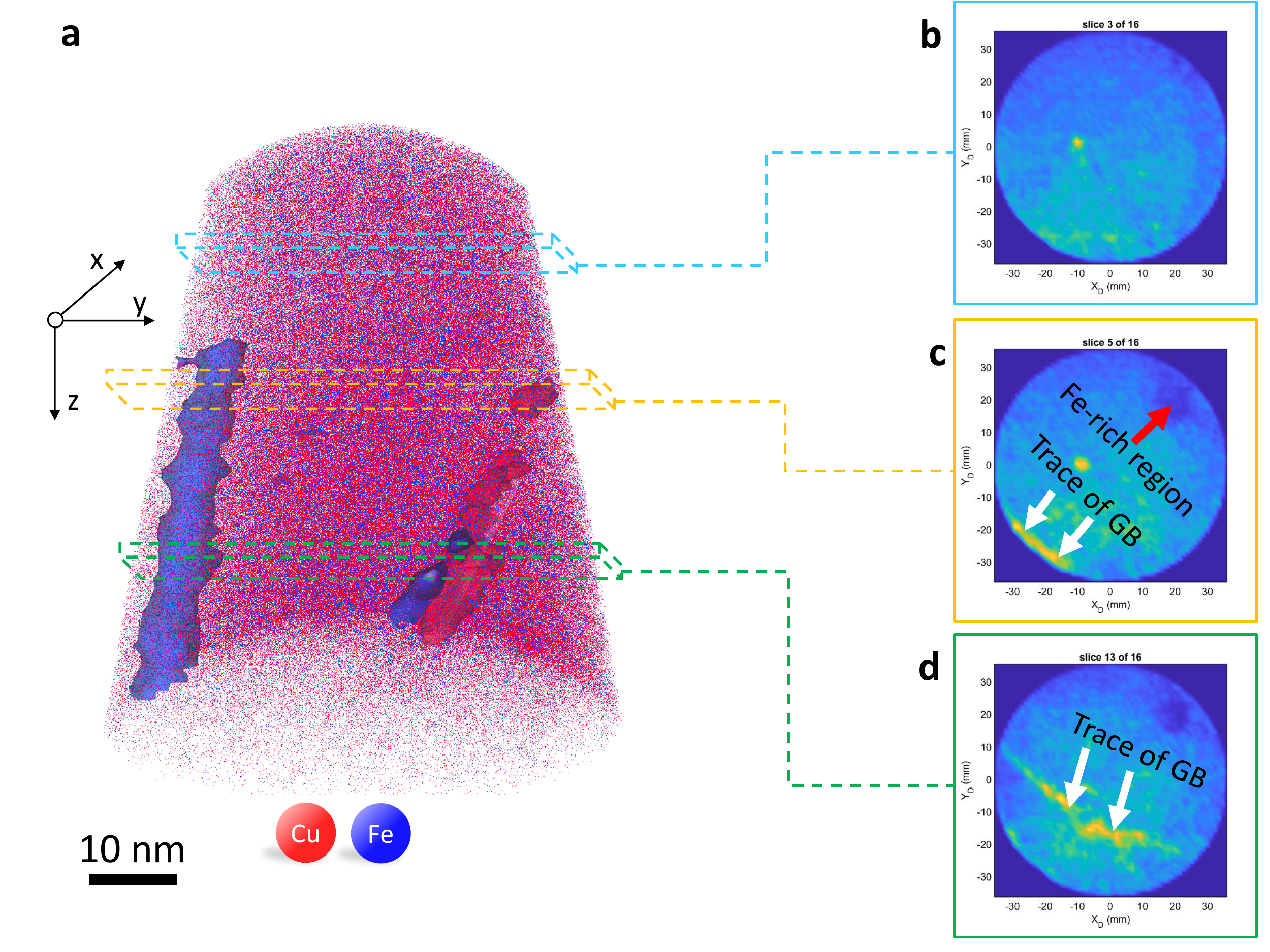}
\caption{APT reconstruction of the as-deformed state. The reconstructed volume (a) consists of Fe19.1at.\%-Cu80.8at.\%. Isosurfaces of 28.7at.\% of Fe and 63.7 at.\% of Cu are displayed. Slices of the tomographic reconstruction are used to identify the position of the grain boundary. The distinct pole in (b) corresponds to a strong orientation preference and therefore the presence of a grain. In (c), apart from the pole, a diffuse pattern can be identified, indicating a grain boundary, which is also visible in (d).}
\label{fig:apt2_2d}
\end{figure*}

\textcolor{black}{
In Fig.~\ref{fig:apt2_2d}, the APT reconstruction of Cu-14Fe (wt.\%) in the as-deformed state is shown, which is used to investigate the grain boundary position. The sequence of 2D-projection slices shows either distinct Cu-poles or a diffuse pattern. The Cu-poles can be attributed to a strong atomic ordering in fcc-configuration and therefore to the presence of grains. The diffuse patterns point at the presence of a grain boundary.}

\bibliography{acta_arxiv}

\end{document}